\documentclass[conference]{IEEEtran}
\IEEEoverridecommandlockouts
\usepackage{cite}
\usepackage{amsmath,amssymb,amsfonts}
\usepackage{graphicx}
\usepackage{textcomp}

\usepackage{booktabs} % For formal tables
\usepackage{arydshln}
\usepackage{subcaption,verbatim,engord,enumitem,multirow,bm,caption}
\usepackage{algorithm,algorithmic}
\usepackage{amscd}
\usepackage{mathtools}
\usepackage{balance}

\usepackage{xcolor}
\def\BibTeX{{\rm B\kern-.05em{\sc i\kern-.025em b}\kern-.08em
    T\kern-.1667em\lower.7ex\hbox{E}\kern-.125emX}}
\begin{document}

% AUTHOR: Enter the title, all letters in upper case
\title{Tensor Processing Units for Financial Monte Carlo}

% AUTHOR: Enter the authors of the article, see end of the example document for further examples
\author{Francois Belletti, Davis King, Kun Yang, Roland Nelet, Yusef Shafi, Yi-Fan Chen, John Anderson\\ [12pt]
Google Research\\
Mountain View CA\\
USA\\
belletti@google.com
}

\maketitle

\begin{abstract}
Monte Carlo methods are critical to many routines in quantitative finance such as derivatives pricing, hedging and risk metrics.
Unfortunately, Monte Carlo methods are very computationally expensive when it comes to running simulations in high-dimensional state spaces where they are still a method of choice in the financial industry.
Recently, Tensor Processing Units (TPUs) have provided considerable speedups and decreased the cost of running Stochastic Gradient Descent (SGD) in Deep Learning.
After highlighting computational similarities between training neural networks with SGD and simulating stochastic processes, we ask in the present paper whether TPUs are accurate, fast and simple enough to use for financial Monte Carlo.
Through a theoretical reminder of the key properties of such methods and thorough empirical experiments we examine the fitness of TPUs for option pricing, hedging and risk metrics computation.
In particular we demonstrate that, in spite of the use of mixed precision, TPUs still provide accurate estimators which are fast to compute when compared to GPUs. 
We also show that the Tensorflow programming model for TPUs is elegant, expressive and simplifies automated differentiation.
\end{abstract}


\section{INTRODUCTION}
The machine learning community has developed several technologies to speed up Stochastic Gradient Descent algorithms for Deep Learning~\cite{goodfellow2016deep}, including new programming paradigms, special-purpose hardware, and linear-algebra computation frameworks. This paper demonstrates that the same techniques can accelerate Monte Carlo integration of stochastic processes for financial applications.

\subsection{Monte Carlo estimation in finance and insurance}
A key problem when pricing a financial instrument --- for insurance or speculation --- is to estimate an average outcome 
defined by a probability space $\left(\Omega, \mathcal{F}, \mathbb{P} \right)$:
$
    E_{\mathbb{P}} \left[ f(\omega) \right]
$,
where $E$ denotes the expectation.
In the following we first provide a basic introduction to derivatives pricing.
Namely, we focus on estimating expectations with the Monte Carlo method in cases where random fluctuations are generated by stochastic processes.
We describe how hardware accelerators compute such estimators faster through parallelization.

\subsubsection{Stochastic processes in continuous time}
Stochastic processes remain the main abstraction employed to model financial asset prices.
Consider a filtered probability space 
$
\left( 
    \Omega, \mathcal{F}, \mathbb{F}, \mathbb{P}
\right)
$
(where $\mathbb{F} = \left\{ \mathcal{F}_t \right\}$ is the corresponding canonical filtration)
supporting a $q$ dimensional Brownian motion $W$ and the Stochastic Differential Equation (SDE)
\begin{equation}
\label{eq:SDE}
    \text{d}X_t = \mu(t, X_t)\text{d}t + \sigma(t, X_t)\text{d}W_t, \: t \in \left[0, T\right].
\end{equation}
The drift ($\mu$) and volatility ($\sigma$) functions take values respectively in $\mathbb{R}^p$ and $\mathbb{R}^{p, q}$.
By definition, a strong solution $(X)$ to Eq. (\ref{eq:SDE}) is a process taking values in $\mathbb{R}^p$
such that $\int_{s=0}^T \left(||b(s, X_s)||_2 + ||\sigma(s, X_s) ||^2_2\right) ds$ is almost surely finite and 
$$
    X_t = X_0 + \int_{0}^t b\left(s, X_s \right)\text{d}s + \int_{0}^t \sigma \left(s, X_s \right) \text{d}W_s, \: t \in \left[0, T\right].
$$
Assuming that $X_0$ is a random variable with finite variance independent of $W$, 
and that $|| b(\cdot, 0) ||_2$ and $|| \sigma(\cdot, 0) ||_2$ are square integrable (as functions of $t$), the existence of a finite Lipschitz constant $K$
such that
$$
    ||b(t, x) - b(t, y)||_2 + ||\sigma(t, x) - \sigma(t, y)||_2
    \leq
    K ||x - y||_2
$$
for all $x, y \in \mathbb{R}^p$ and $t \in [0, T]$
guarantees the existence of such a strong solution on $[0, T]$.
More modern models typically introduce jumps or stochastic volatility which adds realism to the corresponding simulations but does not radically change the underlying computation patterns.

\subsubsection{Monte Carlo methods in finance and insurance}
Monte-Carlo methods rely on simulation and numerical integration to estimate 
$
    E_{\mathbb{P}}[f(X_T)] \text{ or } E_{\mathbb{Q}}[f(X_T)]
$
under the historical or risk-neutral probability ($\mathbb{P}$ and $\mathbb{Q}$ respectively)~\cite{glasserman2013monte}.
Some contracts defining financial derivatives may specify a \emph{path dependent} outcome---such as Barrier or Asian options---in which case the theory of Black, Scholes and Merton still leads us to estimate
$
    E_{\mathbb{P}}[f(X_{0:T})] \text{ or } E_{\mathbb{Q}}[f(X_{0:T})]
$
where $X_{0:T}$ denotes the observation of the process $X$ on the interval $[0, T]$.
In general, we thus seek an estimator for an expectation of the type
\begin{equation}
\label{eq:estimated}
    E[f(X_{0:T})] \text{ where } (X_t) \text{ solves (1) on } [0, T].
\end{equation}
Monte-Carlo methods rely on numerical discretization and integration to produce an estimator for (\ref{eq:estimated}) in the form of an empirical mean over simulated trajectories $\left\{ \widetilde{X}^n_{0,T} | i = 1 \dots N \right\}$:
\begin{equation} \label{eq:MC}
    \widehat{I}_N = \frac{1}{N} \sum_{n=1}^N f \left( \widetilde{X}^n_{0,T} \right).
\end{equation}

In general, because the dynamics of $(X_t)$ are specified in continuous time with real values, a computer-based simulation will suffer from bias coming from the limited precision in numerical representations and more importantly the temporal discretization.
The variance of the the Monte Carlo estimator is also a problem: typically 
if it costs $O(N)$ samples to produce a result with a confidence interval of size $1$
then reducing the interval's size to $\epsilon$ comes at a cost of $O(\frac{N}{\epsilon^2})$ simulations.
Such a rate of convergence can be accelerated thanks to Quasi-Random Monte Carlo methods~\cite{pages2018numerical,joe2008constructing,joe2008notes,sobol2011construction} which then enable a near linear rate of convergence.
Unfortunately, the computation time generally scales as $O(q^2)$ (i.e. quadratically in space) when correlations between different components of $(X_t)$ are taken into account.
For these reasons, Monte Carlo methods still constitute some of the most computationally intensive tasks running routinely at scale across the financial industry.
Accelerating Monte Carlo estimation and making it more cost effective has been a long standing challenge for derivative pricing, hedging and risk assessment.

\subsubsection{Greeks and sensitivity analysis}
Monte Carlo methods in quantitative finance are also used to estimate sensitivities of derivatives' prices with respect to model parameters and the current state of the market. Sensitivities to market parameters---the financial ``Greeks''~\cite{glasserman2013monte}---are used not only to quantify risk, but also construct hedges and synthetic replicating portfolios~\cite{pages2018numerical,glasserman2013monte}.
%For example, the ``delta'', the sensitivity of the price of an option with respect to the current price of the underlying(s), specifies the amount of the underlying(s) that needs to be held in a replicating portfolio.
Automated differentiation of Monte Carlo pricers (also known as AAD) is now a solution of choice in quantitative finance as it is more computationally efficient than methods such as bumping 
to compute sensitivities with respect to many different inputs and parameters~\cite{savine2018modern}.
Tensorflow was designed with automated differentiation at its very core as this technique---often referred to as ``Back-Propagation''---is of key importance for training machine learning models by Stochastic Gradient descent~\cite{abadi2016tensorflow}.
Moreover, Tensorflow readily offers the opportunity to accelerate simulations and automated differentiation without requiring any additional code by enabling researchers to leverage modern hardware such as GPUs and TPUs.

\subsection{Contributions}
In the present paper we focus on leveraging Tensor Processing Units (TPUs) with Tensorflow for financial Monte Carlo methods.
We aim to show that although such accelerators were designed primarily to accelerate the training of Deep Learning models by Stochastic Gradient Descent, TPUs provide cutting edge performance for Monte Carlo methods involving discretized multi-variate stochastic processes.
In particular, we present the following contributions:
\begin{itemize}
    \item We demonstrate that, in spite of the limited numerical precision employed natively in matrix multiplications on TPUs, accurate estimates can be obtained in a variety of applications of Monte Carlo methods that are sensitive to numerical precision.
    %In particular we consider numerical experiments involving pricing and sensitivity analysis for exotics as well as Value-at-Risk (VaR) estimation.
    % \item We argue that TPUs offer efficient risk assessment solutions: for risk metrics being able to simulate many more scenarios is of key importance.
    %, while in cases where limited precision could become an issue, Multi-Level Monte Carlo methods can potentially efficiently correct precision-related bias.
    \item We benchmark the speed of TPUs and compare them to GPUs which constitute the main source of acceleration for general purpose Monte Carlo methods outside of Field Programmable Gate Arrays (FPGAs) and Application-specific Integrated Circuits (ASICs) based solutions.
    %We also focus on random number generation and benchmark the throughput of Pseudo-Random-Number-Generators (PRNGs) on TPU, separatly from vectorized computation and matrix-matrix multiples. 
    \item We show that Tensorflow~\cite{abadi2016tensorflow} constitutes a high level, flexible and simple interface that can be used to leverage the computational power of TPUs, while supporting automated differentiation.
\end{itemize}
The present paper demonstrates that Tensorflow constitutes a flexible programming API which enables the implementation of different simulation routines while providing substantial benefits by running computations in the Cloud on TPUs.
A key consequence is that experiences that used to be iterative for developers now become interactive and inherently scalable without requiring investing in any software or hardware.
We believe such improvements can make financial risk management more cost-effective, flexible and reactive.

\section{RELATED WORK}

\subsection{Pricing techniques and typical computational workloads}
As a first step towards acceleration,
we now examine three typical computational workloads routinely employed to price derivatives and assess risk in quantitative finance.

\subsubsection{SIMD element-wise scalar ops in Euler-Maruyama discretization schemes}
A first characteristic computational workload is associated with mono-variate geometric Brownian models and their extensions in the form of local~\cite{dupire1994pricing,guyon2011smile} or stochastic volatility models~\cite{gatheral2014volatility,bergomi2015stochastic}.
The Euler-Maruyama scheme discretizes SDE~(\ref{eq:SDE}) explicitly forward in time.
Consider the simulation of $N$ independent trajectories,
$$
    \widetilde{X}^n_{t_{i + 1}} =
    \widetilde{X}^n_{t_i} +
    \mu \left(t_i, \widetilde{X}^n_{t_i} \right) \Delta_{t_i} +
    \sigma \left(t_i, \widetilde{X}^n_{t_i} \right) \sqrt{\Delta_{t_i}} Z^n_{i+1}
$$
for $n=1,\dots,N$
where $\widetilde{X}^n_{t_0}=X_0 \in \mathbb{R}$, $\Delta_{t_i} = t_{i + 1} - t_i$, $Z^n_{i+1}$ are Pseudo-Random Numbers distributed following $\mathcal{N}(0, 1)$.
% In the uni-variate case, where the process being simulated has a single scalar component, 
Simulations then reduce to scalar add/multiplies which are independent across simulated scenarios
and is therefore embarrassingly parallel as a clear example of a Single Instruction Multiple Data (SIMD) setting where the different elements of the data undergo independent computations.
Such simulations are trivial to parallelize across simulated scenarios provided Pseudo or Quasi-Random Numbers (PRNs and QRNs respectively) can be correctly generated in parallel~\cite{coddington1997random,thomas2009comparison,salmon2011parallel,l2017random,joe2008constructing,joe2008notes,sobol2011construction}.
An averaging reduction across samples concludes the task.

\subsubsection{Matrix-multiply ops in Multi-variate simulations of correlated processes}
The Euler-Maruyama discretization scheme for scalar stochastic processes naturally extends to the multi-variate setting where each stochastic process takes values in $\widetilde{X}^n_{t_i} \in \mathbb{R}^p$.
However, a major computational difference arises.
If the underlying Brownian motion is in $\mathbb{R}^q$, each simulated time-step in each scenario will require calculating $\sqrt{t_{i + 1} - t_i} \sigma \left(t_i, \widetilde{X}^n_{t_i} \right) Z^n_{i+1}$
with $Z^n_{i+1} \sim \mathcal{N}(0, I_q)$ and $\sigma \left(t_i, \widetilde{X}^n_{t_i} \right) \in \mathbb{R}^{p, q}$,
which implies that a $p \times q$ matrix/vector product has to be computed.
If $N$ scenarios are stacked together to benefit from the corresponding hardware acceleration, the operation becomes a $p \times q, q \times N$ matrix/matrix products.
Here as well, a final reduction averages the simulated outcomes.

\subsubsection{Chained linear system inversions in the Longstaff-Schwartz Method (LSM) for value estimation}\label{LSM}
The regression-based estimation method proposed by Longstaff and Schwartz~\cite{longstaff2001valuing} to price American Options has become a standard pricing method for callable financial instruments (e.g., American or Bermuda options) with high dimensional underlyings (e.g., call on a maximum or weighted combination of stocks).
In the setting of callable options, which can be exercised at multiple points in time before their maturity,
the pricing problem is solved using a Monte-Carlo method to simulate future trajectories as well as a dynamic programming approach to back-track optimal decisions in time.
To enable dynamic programming, for each decision instant $t_i$, a classic approach is to estimate a Value Function on the state of the underlying:
$$
    X_{t_i} \mapsto V_{t_i}(X_{t_i}) = \text{max} \left(f(X_{t_i}), E(V_{t_{i+1}} | X_{t_i}) \right)
$$
with the convention that $V_T = f(X_T)$ where $f$ is the option's payoff function.
As the conditional expectation $E(V_{t_{i+1}} | X_{t_i})$ is the closest square integrable random variable to $X_{t_i}$ (according to the $L_2$ norm) 
LSM fits a model to interpolate $E(V_{t_{i+1}} | X_{t_i})$ between values of $X_{t_i}$ that have actually been simulated.
LSM employs a linear-regression on a set of $K$ features derived from the the simulated values of $X_{t_i}$ such as $(1, X_{t_i}, X^2_{t_i}, \dots)$ or a finite number of Hermite polynomials evaluated at the simulated values~\cite{glasserman2013monte,longstaff2001valuing}.
Given a set of simulated values $\left\{ \widetilde{X}^n_{t_i} | n = 1 \dots N\right\}$,
the set of values $\left\{V\left(\widetilde{X}^n_{t_i}\right) | n = 1 \dots N\right\}$
is projected onto the set of regressors
$\left\{
    \psi_1\left(
        \widetilde{X}^n_{t_i}
    \right), 
    \dots,
    \psi_K\left(
        \widetilde{X}^n_{t_i}
    \right), 
    | n = 1 \dots N\right\}
$
where $\left(\psi_k\right)_{k=1 \dots K}$ are featurizing functions (e.g., Hermite polynomials).
Therefore, a linear regression of $N$ scalar observations is needed, for each candidate time-step to exercise the option, onto $N$ vectors of $K$ dimensions.
Typically, for efficiency, a Cholesky decomposition of the Grammian will be computed to effectively solve the linear regression.
This computational cost adds to the cost of simulating the original paths for the trajectory of the underlying asset(s) which may themselves be correlated.
The overall procedure yields a price estimate as the expected value function at the first exercise time:
$
    \widehat{I}_N = \frac{1}{N} \sum_{n=1}^N V \left( \widetilde{X}^n_{t_0} \right).
$
Therefore, an averaging reduction also concludes LSM.

\subsection{Pre-existing hardware acceleration strategies}
Having reviewed a few algorithms that are core to financial Monte Carlo, we now give an overview of hardware-based techniques to reduce their running time.
A first approach to accelerate Monte Carlo methods had consisted in running them on High Performance Computing (HPC) CPU grids with parallelization paradigms such as Message Passing Interface (MPI).
We focus here on device-level acceleration with hardware accelerators that can be used as elements of a distributed compute grid if needed.

\subsubsection{GPUs}
The rise of general purpose high level APIs for scientific calculations on GPUs with CUDA or OpenCL has prompted a wide development of GPU-based approaches to accelerate Monte Carlo methods. Pricing and estimating risk metrics in finance 
are especially well-suited to acceleration by GPUs, due to the embarrassingly parallel nature of Monte Carlo Methods, and to their use of computationally intensive linear algebra routines.
% has been an application of choice for such hardware due to the embarrassingly parallel nature of Monte Carlo methods and the reliance of simulations on computationally intensive linear algebra routines.
Generating PRNs and QRNs in parallel correctly~\cite{coddington1997random,thomas2009comparison,salmon2011parallel,l2017random,joe2008constructing},
coupled with algorithmic re-factorization to fully utilize GPUs have enabled large speedups with respect to CPUs for pricing~\cite{lee2010utility,podlozhnyuk2008monte,abbas2014pricing,marshall2011simulation}, risk metrics~\cite{dixon2012monte} and sensitivity analysis~\cite{du2013adjoint}.

\subsubsection{FPGAs}
Many works have demonstrated that Field Programmable Gate Arrays (FPGAs) provide substantial speedups with respect to GPU implementations and reduce energy costs in servers. 
While some methods have employed FPGAs as standalone solutions~\cite{weston2012rapid,weston2010accelerating},
other approaches have used a mixed precision approach relying on both a CPU and an FPGA~\cite{chow2012mixed,brugger2014hyper}.
In particular, Multi-Level Monte Carlo (MLMC)~\cite{giles2015multilevel} can be applied to FPGAs computing low resolution fast simulations paired with CPUs running at reference precision.

\section{TENSOR PROCESSING UNITS}
A Tensor Processing Unit (``Cloud TPU'' or ``TPU'')---a custom-developed application-specific integrated circuit (ASIC) specialized for deep learning---offers $420\times 10^{12}$ floating-point operations per second (FLOPS) and 128GB of high bandwidth memory (HBM) in its latest release. 
%The TPU architecture is abstracted behind the Tensorflow framework and can be deployed both for training large deep neural networks and for performing low-latency online prediction without any knowledge about its native environment.
The TPU architecture is abstracted behind the Tensorflow framework. High-level Tensorflow programs, written without using detailed knowledge of TPUs,  can be deployed on TPU hardware in the cloud to train or serve deep neural networks.
\cite{jouppi2017tpuperf} reports impressive acceleration of training and low-latency online prediction.

Although TPU targets deep learning, it is flexible enough to address computational challenges in various fields. In the present paper, we particularize its application to financial Monte Carlo.

\subsection{TPU System Architecture}
\begin{figure}[ht]
    \centering
    \includegraphics[width=0.95\linewidth]{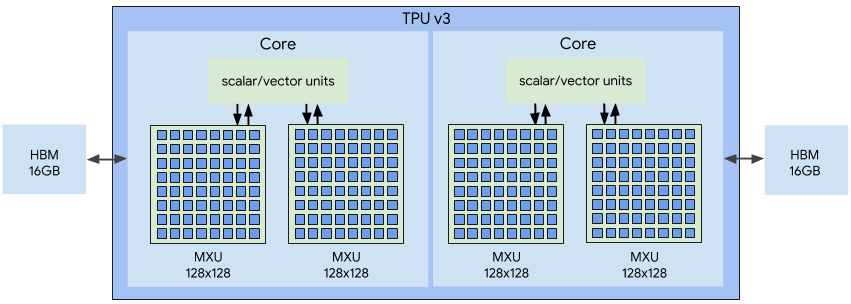}
    \includegraphics[width=0.6\linewidth]{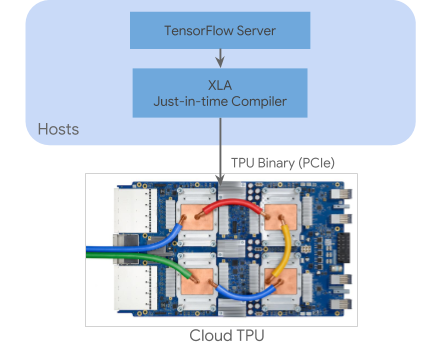}
    \caption{Hardware architecture and programming model of Tensor Processing Units (TPUs)~\cite{cloudtpu}.}
    \label{fig:TPU}
\end{figure}

One TPU is comprised of four independent chips. Each chip consists of two compute cores called Tensor Cores. A Tensor Core, as shown in Fig. \ref{fig:TPU} consists of scalar, vector and matrix units (MXU). In addition, 16 GB of on-chip High Bandwidth Memory (HBM) is associated with each Tensor Core for Cloud TPU v3 --- its latest generation. 
Communication between Tensor Cores occurs through high-bandwidth interconnects. 
All computing units in each Tensor Core are optimized to perform vectorized operations.
In fact, the main horsepower of a TPU is provided by the MXU which is capable of performing $128\times 128$ multiply-accumulate operations in each cycle~\cite{gcp2019tpuperf}. 
While its inputs and outputs are 32-bit floating point values, the MXU typically performs multiplications at the reduced precision of bfloat16 --- a 16-bit floating point representation that provides better training and model accuracy than the IEEE half-precision representation for deep learning as it allocates more bits to the exponent and less to the mantissa.

\subsection{Programming Model}
Programming for TPUs is generally done through high-level Tensorflow API. When the program is run, a TensorFlow computation graph is generated and sent to the Cloud TPU over gRPC~\cite{cloudtpu}. The Cloud TPU server compiles the computation graph just in time, partitions the graph into portions that can run on a Cloud TPU and those that must run on a CPU and generates Accelerated Linear Algebra (XLA) operations corresponding to the sub-graph that is to run on Cloud TPU. Next, the XLA compiler takes over and converts High Level Optimizer (HLO) operations that are produced by the TensorFlow server to binary code that can be run on Cloud TPU, including orchestration of data from on-chip memory to hardware execution units and inter-chip communication. Finally, the binary is sent to the Cloud TPU for execution.

\subsection{From DNN inference to stochastic process simulation}
In the present paper, we leverage the computational similarity between DNN inference and simulating high dimensional financial portfolios.
Tensorflow and TPUs have been designed to provide a high level interface for deep learning programming to train mission critical models rapidly.
Our proposal is to leverage such performance for a different purpose: running stochastic process simulation for financial applications.

\begin{figure}
    \centering
    \includegraphics[width=\linewidth]{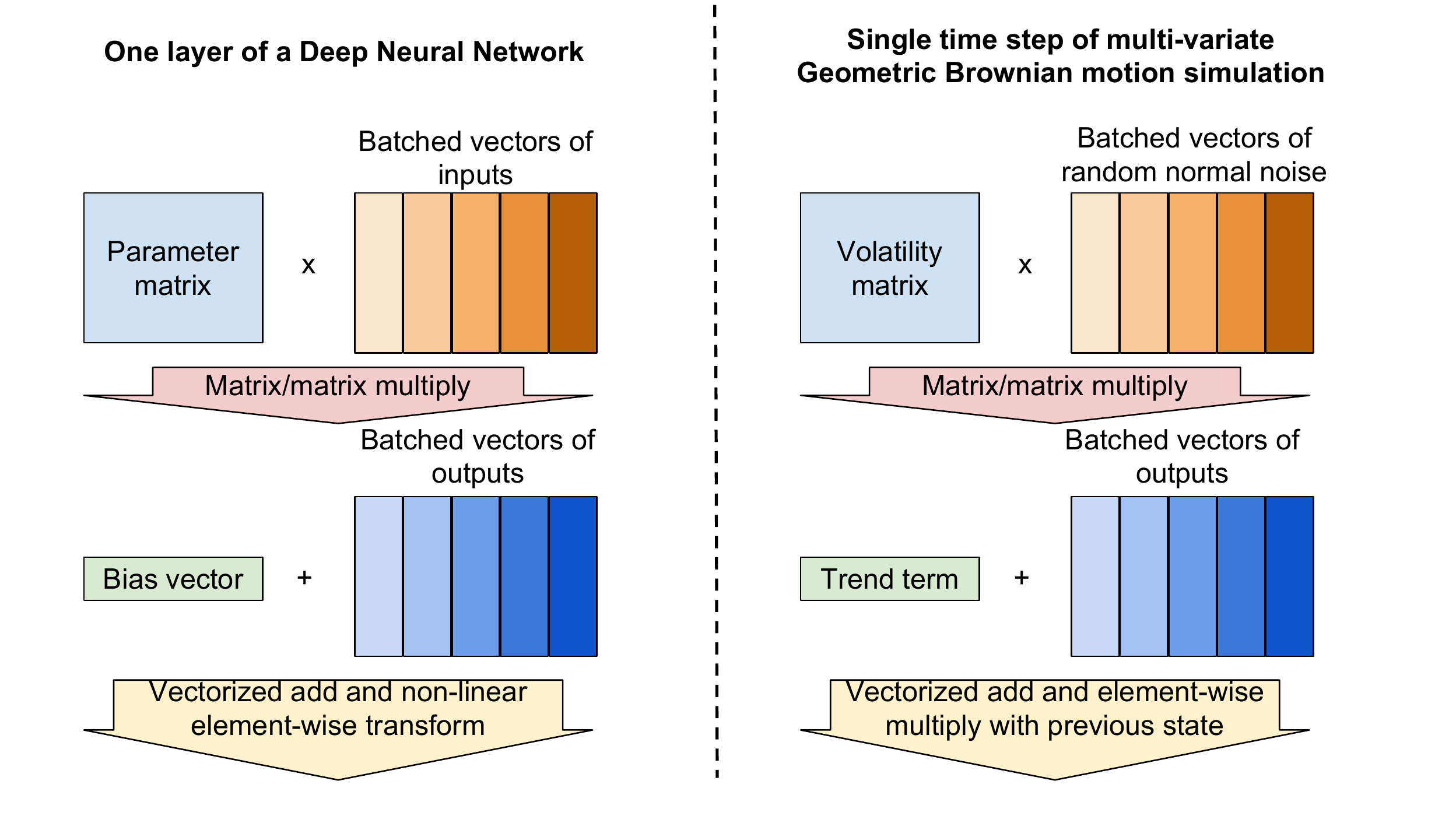}
    \caption{Computational similarity between a DNN layer and a time step of a multi-dimensional Geometric Brownian Monte Carlo.}
    \label{fig:computational_similarity}
\end{figure}

Computing the output of a DNN implies chaining matrix/matrix multiplies interleaved with element-wise vectorizable operations.
Very similar computational patterns are involved in quantitative finance as multi-dimensional simulations
require the computation of matrix/matrix multiples (with batched PRNs) and element-wise vectorizable operations.
Figure~\ref{fig:computational_similarity} illustrates this high computational similarity.
Furthermore, as training DNNs requires computing a gradient through automated differentiation between the outputs and all the parameters of the network, a DNN learning framework such as Tensorflow is ideal to enable sensitivity estimation through path-wise differentiation~\cite{savine2018modern}.

\section{PSEUDO AND QUASI RANDOM NUMBER GENERATORS on TPU}
We describe how Pseudo-Random Number Generators (PRNGs) and Quasi-Random Number Generators (QRNGs) run on TPUs. 

\subsection{PRNGs designed for parallelization and statistical soundness}
For financial Monte-Carlo, the Mersenne Twister 19337 (MT19337)~\cite{matsumoto1998mersenne} constitutes a very popular approach and yields a sequence of PRNs with period $2^{19337}$.
Such a PRN sequence is produced as a single stream and therefore parallelization relies on sub-stream approaches~\cite{savine2018modern,pages2018numerical}
which typically requires deciding ahead of time how many PRNs will be consumed by each core through the simulation.
% Sub-stream parallelization will either generate large non-overlapping sections of the sequence on each core or generate numbers with a skip-ahead to guarantee that distinct elements of the sequence are generated by each processor.
% While sub-stream parallelization produces correct PRN sequences, it has high computational costs, and it 

A more flexible approach for distributed and parallel computing is to use multi-stream random number generation as in~\cite{salmon2011parallel}.
Multi-stream algorithms, such as  Threefry and Philox, use cryptographic methods to generate multiple sequences $(x_n^k)_{n = 1 \dots N, k = 1 \dots K}$ guaranteed to be seemingly statistically independent if each core employs a distinct key.
The number of admissible keys is typically $2^{64}$ if $64$ bit integers are used to encode keys and the period of each sequence is $2^{64}$ if $64$ bits are used to represent the sequence iterator.
\cite{salmon2011parallel} shows that these methods have higher throughput and better statistical properties when submitted to Big-Crunch~\cite{l2007testu01} than standard generators.
In particular, while Threefry and Philox pass all the tests in the battery of Big-Crunch, MT19937 fails most of them.
We have confirmed that the generation of uniformly distributed 32 bit integers on TPU successfully passes Big-Crunch~\cite{l2007testu01} (the double precision version of Big-Crunch is not directly relevant to TPUs which generate uniformly distributed floats in $[0, 1)$ in single precision only).
For this reason and because of the ease of parallelization, Tensorflow employs keyed multi-stream PRNGs. In particular, we employ Threefry in our TPU experiments and Philox on GPU.

\subsection{QRNGs and Sobol sequence on TPU}
While PRNGs aim at generating iid samples, QRNGs purposely generate non-sequentially-independent samples to cover the $[0, 1)^q$ hyper-cube as uniformly as possible and accelerate numerical integration~\cite{pages2018numerical,glasserman2013monte}.
QRNGs are preferred to numerically estimate expectations long as the sampled space is not problematically high dimensional. 
Indeed, while Monte Carlo integration provides a rate of convergence of $O(\frac{1}{\sqrt(N)})$ (where $N$ is the number of samples),
low discrepancy sequences (e.g. the Sobol sequence) outputted by QRNGs can decrease the mean square error at a speed if $O(\frac{\ln(N)^{q-1}}{N})$~\cite{korn2010monte}.
The main shortcoming of low discrepancy sequences is the loss of confidence intervals estimates.
However, randomized Quasi Monte Carlo provides confidence interval estimates
through the use of scrambling or other randomization techniques~\cite{pages2018numerical,glasserman2013monte,l2000variance,owen2003quasi}.

%It is well known that QMC's efficacy tends to decrease in higher dimensions.
%Techniques such as hybrid QMC~\cite{asmussen2007stochastic} help mitigate these issues by using QMC exclusively for the dimensions of highest sensitivity of the simulation while using PRNG based standard Monte-Carlo for the remaining dimensions. 

In the present article, as we want to examine numerical bias with respect to temporal discretization bias and simulation variance, we present our first set of results based on classic Monte Carlo simulations.
We demonstrate empirically in~\ref{sec:sobol} that QMC can run on TPU with no significant overhead as compared to Monte Carlo while benefiting from a near linear rate of convergence (in terms of number of samples).

\section{Numerical precision on TPU and Multi-Level Monte Carlo}
Running dynamical system simulations, in particular in finance, often relies on high numerical precision to produce faithful results.
We now describe the numerical precision of TPUs.

\subsection{Single and bfloat16 precision on TPU}
As opposed to today's CPUs or GPUs, TPUs do not offer double ($64$ bit) precision for floating point representations.
The default representation is single ($32$ bit) precision and scalar multiplications in the MXU for matrix multiplies are computed in bfloat16 ($16$ bit) precision prior to being accumulated in single precision.
As $16$ bits are quite few to represent numbers for ML applications and chained multiplications in general, a non standard representation scheme has been employed to better capture high magnitude values and prevent overflow.

The single precision IEEE floating point standard allocates bits as follows: 1 sign bit, 8 exponent bits and 23 mantissa bits.
The IEEE half precision standard uses 1 sign bit, 5 exponent bits and 10 mantissa bits.
In contrast, the bfloat16 format employs 1 sign bit, 8 exponent bits and 7 mantissa bits.

We found in our numerical experiments that both the single precision --- used in accumulators and vector units for element-wise operations --- and the bfloat16 precision did not yield significantly different results as compared to the double precision in the context of financial Monte Carlo.

\subsection{Numerical precision, discretization bias and variance in risk metrics}
Financial Monte-Carlo methods are typically concerned with the simulation of a continuous time SDE such as Equation (\ref{eq:SDE}).
Analyzing the convergence of the Monte Carlo estimator $\widehat{I}_N$ in Equation (\ref{eq:MC}) hinges upon the well known bias-variance decomposition of the $L_2$ error~\cite{pages2018numerical}:
{
\small
$$
    ||E\left[f(X_{0:T})\right] - \widehat{I}_N||^2_2 = \left(
        E \left[ f(X_{0:T}) \right] - E \left[ \widehat{I}_N \right] 
    \right)^2 + \frac{\text{Var} \left(\widehat{I}_N\right)}{N}. 
$$
}

The bias term typically conflates the biases induced by temporal discretization and floating point numerical representation.

When simulating a SDE, a temporal discretization occurs that induces most of the bias affecting the Monte Carlo simulation.
Indeed, as opposed to the actual process of interest $(X_t)_{t \in [0, T]}$,
the Monte Carlo simulation typically employs a piece-wise continuous approximation $(\widetilde{X}_t)_{t \in [0, T]}$
for which only $H$ values are computed (if $H$ is the number of temporal discretization steps) every $\Delta_t$ (where $\Delta_t$ is the temporal discretization step).
Under several assumptions which are typically true for financial pricing, the Taley-Tubaro theorem~\cite{pages2018numerical} states that the discretization bias reduces as $O(\frac{1}{H})$, that is inversely to the number of steps employed when discretizing the process.
Such a bias term dominates in practice as we will demonstrate numerically. In comparison, the numerical precision induced bias is negligible.
Finally, as in~\cite{chow2012mixed,brugger2014hyper},
the MLMC method~\cite{giles2008multilevel,giles2015multilevel} can efficiently estimate the numerical bias term if it is significant and compensate for it.
In particular, MLMC helps low precision fast implementations accelerate reference precision Monte Carlo by running them as a companion estimator whose bias can be identified with a few samples and is practically employed to lower the variance of the Monte Carlo estimator by generating many samples quickly.

\section{NUMERICAL EXPERIMENTS}
We now demonstrate that TPUs are fast instruments whose numerical precision is commensurate with the needs of financial Monte Carlo based risk metrics.
In the following, we use the Philox on GPU and Threefry on TPU (where it is more challenging to support certain low-level operations leveraged by Philox).
Also, we use the same python Tensorflow code on GPU and TPU for two reasons:
we want to make our comparison with a library that has been highly optimized for speed and we want to make comparisons of speed at equal level of software engineering effort.
Here NVidia v100 GPUs are used to anchor our results, but no architecture specific optimizations have been done, which could considerably improve their performance.
Therefore, the GPU wall times we give can only be considered a solid reference but not an indicator of the peak performance one can obtain on the NVidia V100 GPU.
In the following, we do not take into account compilation times when measuring wall time on TPU as the corresponding overhead is obviously amortized in most applications that are latency sensitive, and is not a hindrance for an interactive experience in a Colaboratory notebook (interactive notebooks comparable to ipython Jupyter notebooks).

\subsection{European option pricing and hedging}
Our first experiments are concerned with European option pricing, i.e. non-callable derivatives with a single terminal payoff.

\subsubsection{Uni-variate process simulation to benchmark TPUs' VPU}
\begin{figure}[ht]
    \centering
    \includegraphics[height=1.28\linewidth]{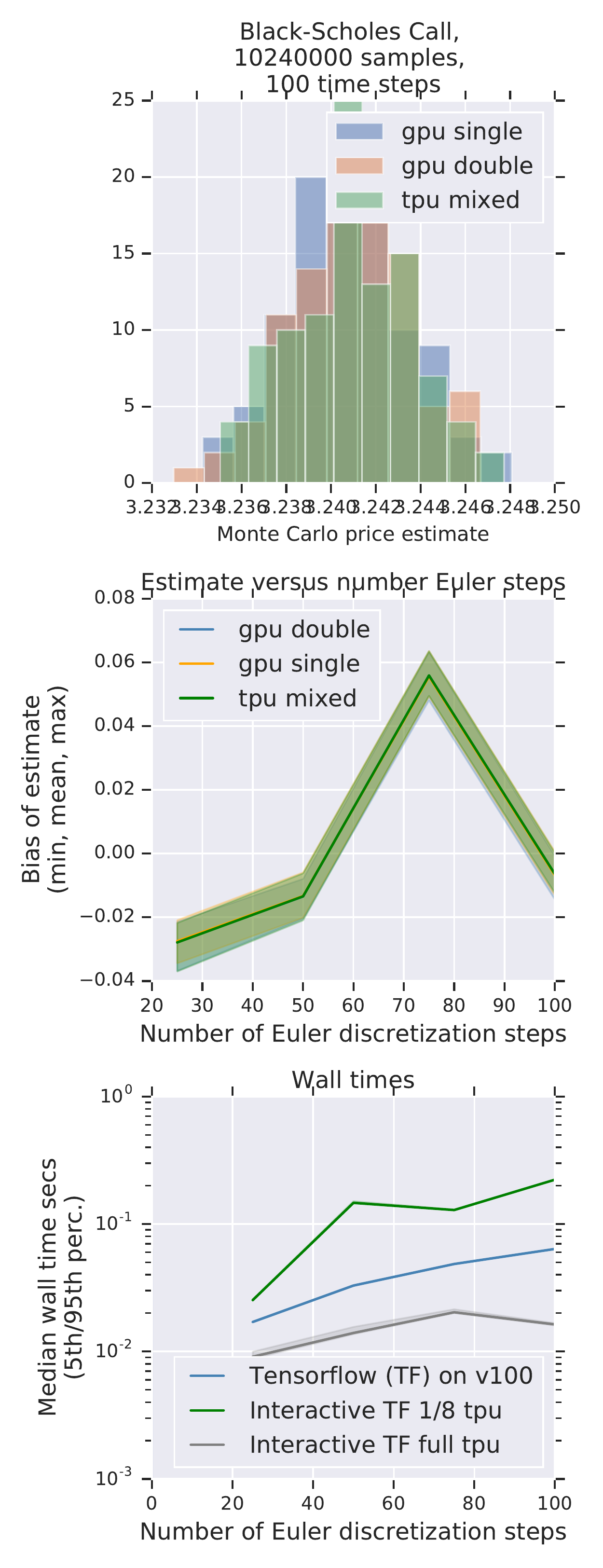}
    \includegraphics[height=1.28\linewidth]{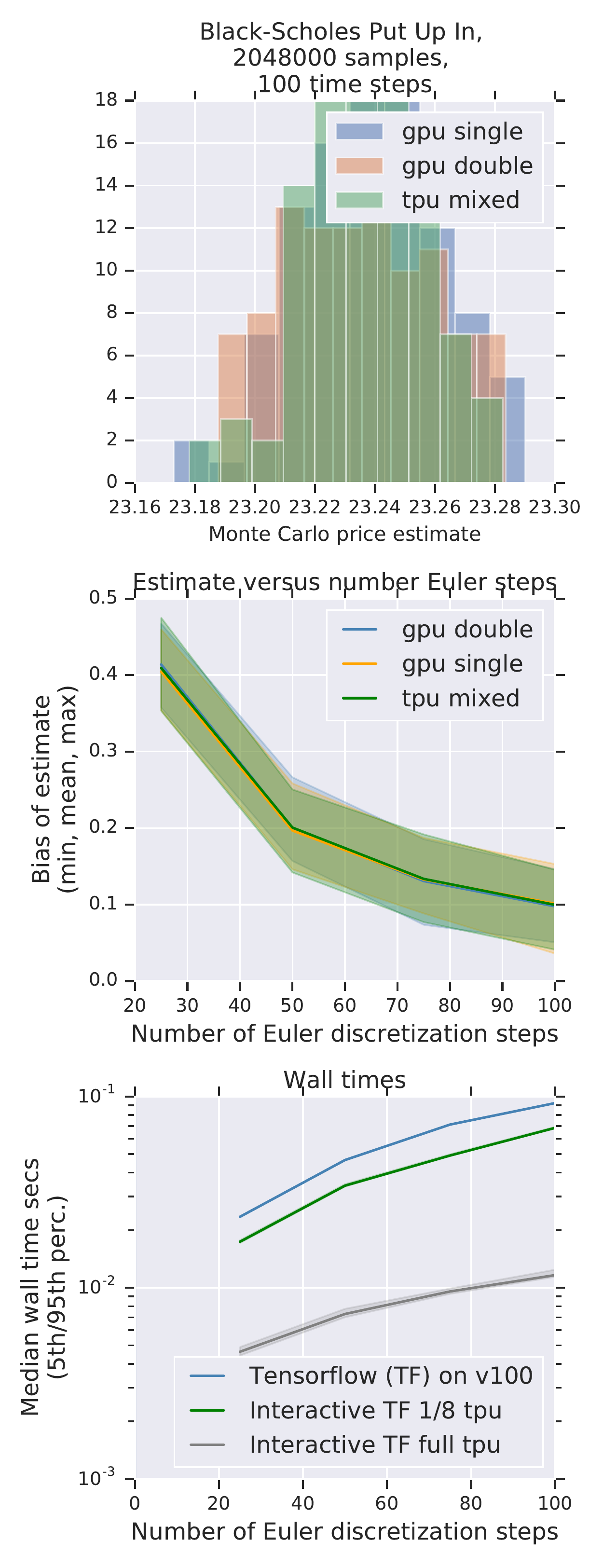}
    \caption{
    \textbf{Vanilla call (left) and Path dependent exotic Put (right):} Estimate distributions in single and double precision. Bias terms with various numerical precision and discretization steps for the estimation of the payoff's risk neutral expectation under a Black-Scholes model help assess the impact of the temporal discretization step size. The temporal discretization bias clearly dominates the numerical precision bias.
    }
    \label{fig:univariate_vanilla}
    \label{fig:univariate_exotic}
\end{figure}

Uni-variate stochastic process simulations do not need matrix multiplies and therefore constitute helpful benchmarks to specifically assess the performance of the Vector Processing Unit (VPU) on TPUs,
which performs single (32-bit) floating point arithmetic.
\\
\textbf{Vanilla Call:}
First we start with the extremely simple example of pricing a $1$ year maturity European Call option (strike at $120$) under the standard Black-Scholes model with constant drift ($0.05$) and volatility ($0.2$) with an initial underlying price of $100$.
The analytic price of the option in double precision is $3.24747741656$.
What we intend to show here --- as we know the actual option price analytically --- is that temporal discretization bias largely dominates over numerical precision bias.
Each of the $100$ simulation runs has $25$ to $100$ discretization steps and $10$M samples.
One can verify in Figure~\ref{fig:univariate_vanilla} that TPUs provide a comfortable speed-up compared to GPUs running the same Tensorflow graph, and that the bias induced by the use of lower precision is negligible compared to that induced by the temporal discretization.
\\
\textbf{Path dependent exotic Put:}
We make things slightly more complex and consider pricing an In and Out Put (activating Barrier).
The initial price of the underlying is $100$, the strike $120$, the barrier $140$.
The drift is still constant ($0.03$) as well as the volatility ($0.8$).
The analytic price of the option (given by the symmetry principle under the Black-Scholes model) in double precision is $23.1371783926$.
Between each two discretization steps with values $x$ and $y$ we simulate the maximum of the Brownian bridge --- classically~\cite{glasserman2013monte,pages2018numerical} with 
$m \sim 0.5 \left(x + y + \sqrt{\left( \left(x - y \right)^2 \right) - 2 \Delta t \sigma^2 \log(U)} \right)$ where $U$ is distributed uniformly in $(0, 1)$
---
to reduce the discretization bias of the method.
Such an experiment also demonstrates that programming in Tensorflow is flexible enough to allow for varied simulation schemes.
Each of the $100$ simulation runs has $25$ to $100$ discretization steps and $2$M samples.
Again, in Figure~\ref{fig:univariate_exotic}, one can appreciate the speed up for a Tensorflow graph running on TPU with respect to an identical graph running on GPU while the impact of the use of single precision as opposed to double precision is hardly noticeable.

% \emph{Conclusion on the uni-variate experiments:}
% Simulating more steps and more samples is more instrumental to risk assessment than increasing the precision of the floating point representation being used.
% From an interactive python front-end (a Google Colaboraty notebook), it takes as low as $10$ milliseconds (in median over $100$ runs) to run the pricing routine (including round-trip time to the TPU).
% Therefore, we argue that TPUs can be used as tools to enable fast iteration for research related to Monte Carlo methods in finance.

\subsubsection{Multi-variate process simulation to benchmark TPUs' MXU}\label{sec:basket_option}
Multi-variate stochastic simulations represent heavier computational workloads than their uni-variate counterparts whenever they involved correlated dynamics.
A matrix/matrix multiply is then involve at each step of the simulation when computing the product of the volatility matrix with the multidimensional normally distributed stacked PRNs.
In such a setting, the speed of the MXU on TPUs may be beneficial, but, as it uses a custom floating point representation, one needs to assess that no substantial numerical precision bias appears.

\begin{figure}[ht]
    \centering
    \includegraphics[height=1.28\linewidth]{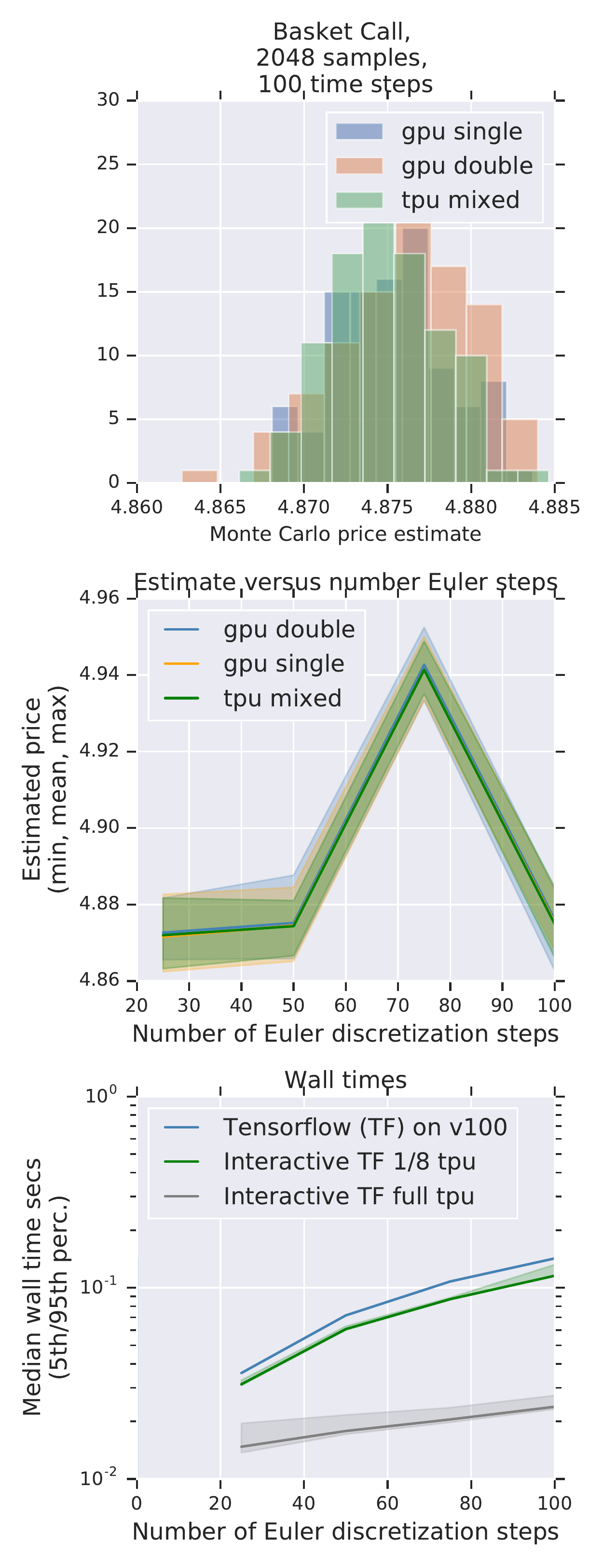}
    \includegraphics[height=1.28\linewidth]{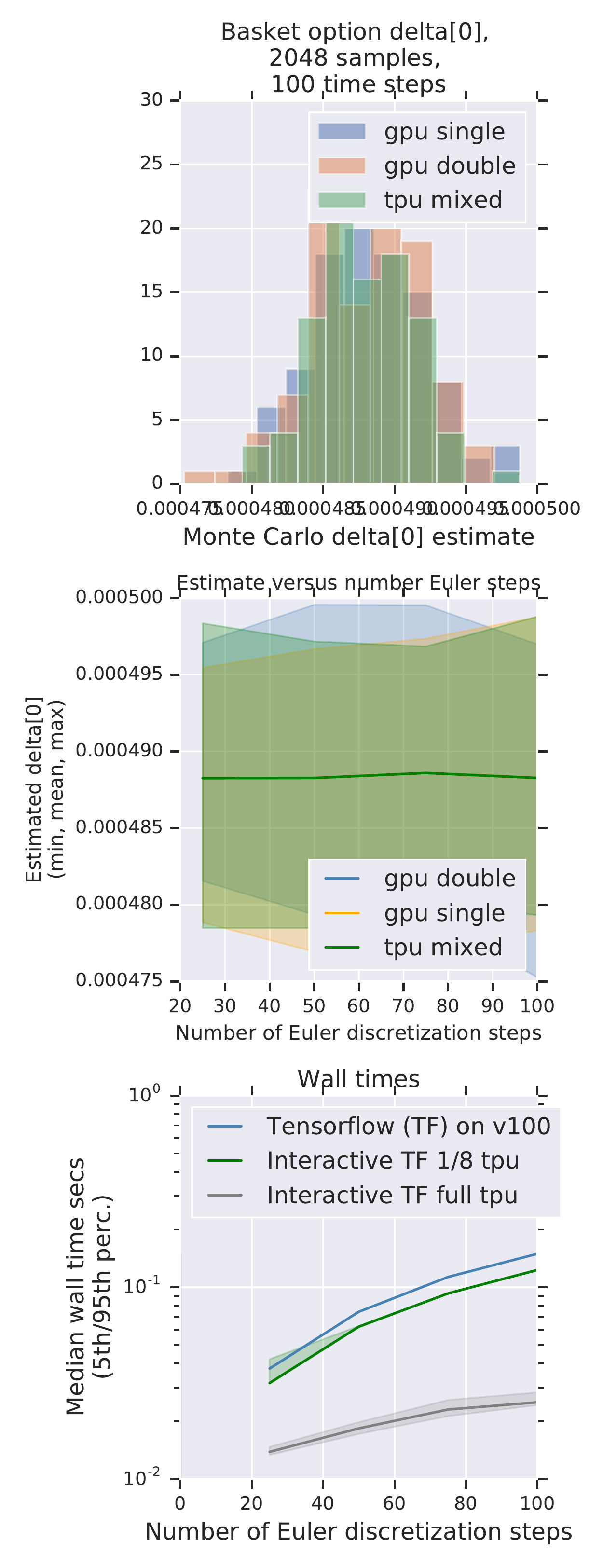}
    \caption{\textbf{Basket European Option (left), first component of Basket European Option Delta (right) :} Estimate distributions and wall times when the MXU is employed on TPU with mixed precision. No significant bias is introduced by running computations on a TPU. The measured wall times, which include the network round trip between front end host and TPU, are very competitive.
    }
    \label{fig:european_basket_pricing}
    \label{fig:european_basket_pricing_delta}
\end{figure}

\textbf{Basket European option:}
We price an at-the-money Basket European call with $2048$ underlyings whose price is initially $100$.
The interest rate is $0.05$ and the volatility matrix we use is a historical estimate based on market data collected on daily variations of randomly selected stocks from the Russell 3000 through $2018$.
$2$K samples are used for each of the $100$ simulations.
Simulations now involve matrix multiplications corresponding to Equation (\ref{eq:SDE}) and therefore the MXU of the TPU is used with a reduced bfloat16 precision.
All other computations on TPU run in single precision. In Figure~\ref{fig:european_basket_pricing} we present the estimates provided in mixed precision on TPU and compare them with single and double precision estimates.
We find that running simulations on TPU does not introduce any significant bias while offering substantial speed ups compared to GPUs.

\textbf{Basket European option Delta:}
As automated differentiation is integrated into the Tensorflow framework, almost no additional engineering effort is required to compute path-wise sensitivities of a MC simulation.
Considering the same European option, we now compute its ``delta'', that is to say the first order derivative of the option's price estimate with respect to the initial price vector (with $2048$ components).
As demonstrated in the code snippet of Figure~\ref{fig:code_snippet}, presented in section~\ref{sec:programming}, computing such a sensitivity estimate, which can represent a substantial software engineering effort for libraries not designed with AAD in mind~\cite{savine2018modern}, only requires a single line of code in Tensorflow.
In Figure~\ref{fig:european_basket_pricing_delta}, we can appreciate that although back-propagation introduces an additional chain of multiplications, no significant bias is added (we present results here only for the first component of the ``delta'' for ease of reading).

\subsection{Risk metrics}
The impact of mixed precision on variance may become a concern for risk metrics such as VaR and CVaR whose purpose is to estimate percentiles and losses for a given portfolio that occur in rare adverse scenarios.

\subsubsection{Estimating Value-at-Risk with many underlying factors}\label{sec:CVAR}
In this simulation, we consider the simulation of the same $2048$ underlying assets as in the Basket option experiment (all from the Russel 3000) with a trend and correlation structure estimated based on historical data on daily variations.
The portfolio whose loss distribution we sample from consists of 5 call options with different strikes on each of the underlying assets.
As a result the portfolio of interest has $10240$ instruments.

\begin{figure}[ht]
    \centering
    \includegraphics[height=1.28\linewidth]{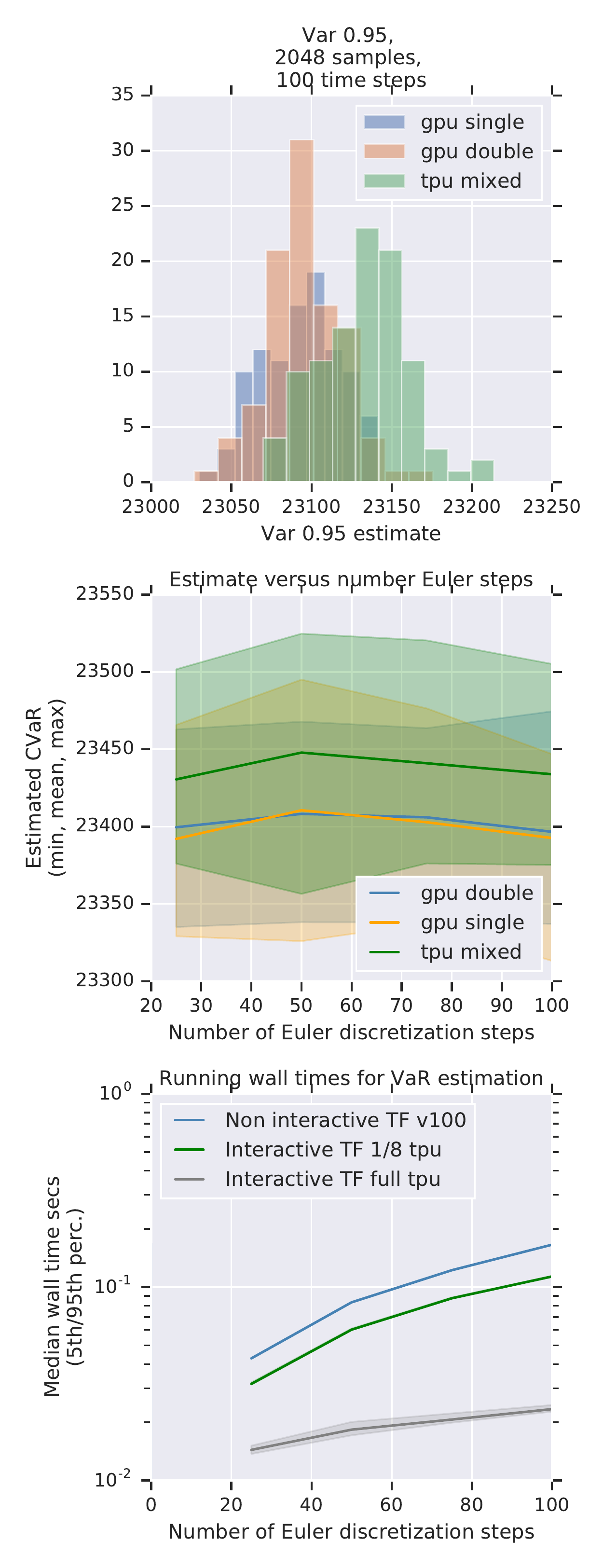}
    \includegraphics[height=1.28\linewidth]{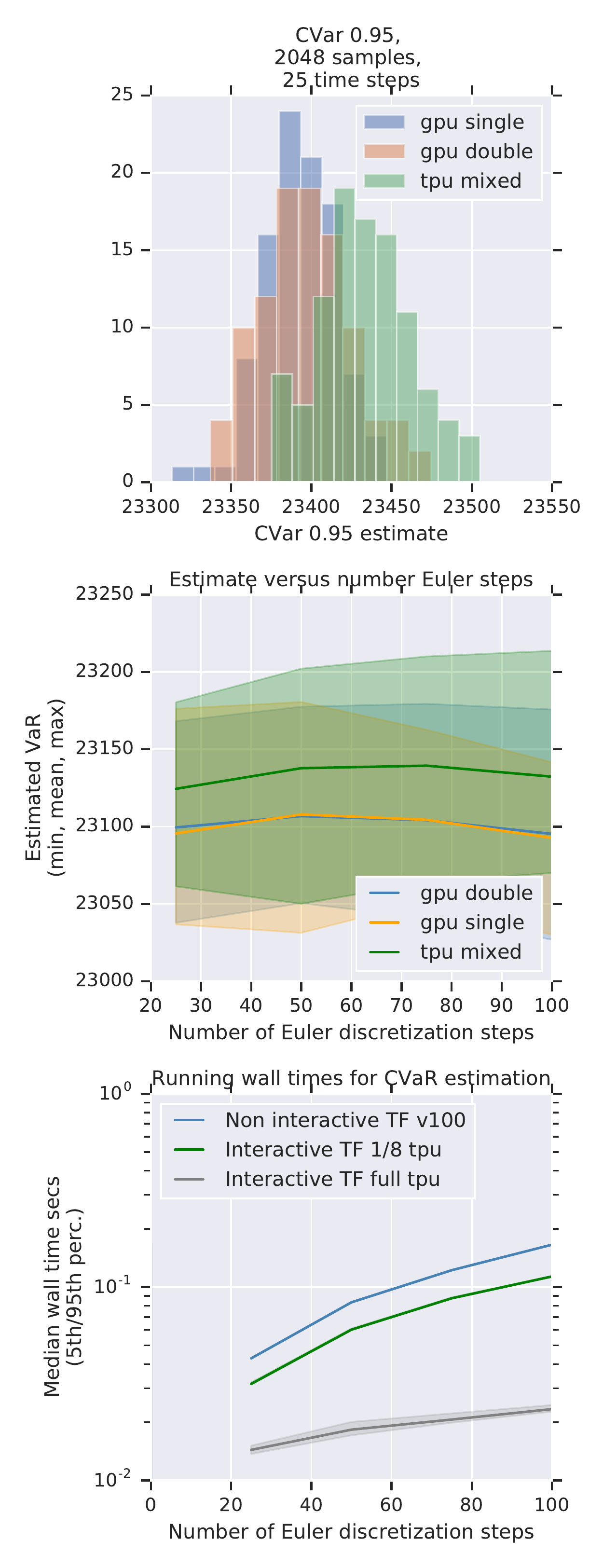}
    \caption{\textbf{$\text{VaR}_{0.95}$ (left) and $\text{CVaR}_{0.95}$ (right):} Estimates of  and corresponding wall times (the MXU is used with bfloat16 precision on TPU for matrix multiplies).
    Little bias is introduced by running computation in mixed precision on TPU. The measured wall times (including the network round trip for TPUs) are very competitive for TPU.
    }
    \label{fig:var}
    \label{fig:cvar}
\end{figure}

\textbf{Value-at-Risk:}
We simulate the distribution of profit and losses (PnL) for the portfolio of interest over the course of a year with different scales of temporal discretization.
The first risk metric of interest is the standard Value-at-Risk (VaR) at level $\alpha$.
By definition, VaR is a quantile of the distribution of losses on an investment in the presence of contingencies.
Given a random variable $\text{PnL}(\omega)$ representing the PnL of the overall portfolio subjected to random perturbations $\omega$,
we have to estimate the quantile of level $\alpha$ of the PnL distribution:
$$
\text{VaR}_{\alpha} \left( \text{PnL}(\omega) \right) = 
    - \text{inf}
    \left\{ 
        x \in \mathbb{R} : \mathbb{P} \left( \text{PnL}(\omega) \geq x \right) > \alpha
    \right\}.
$$
The results presented in Figure~\ref{fig:var} show that limited precision in the MXU has some impact in the estimated VaR terms as little bias is present (less than $1\%$ in relative magnitude compared to a double precision simulation).
However, the speed-ups reported are substantial, so that a MLMC approach could be employed to preserve most of the computational gains while identifying the TPU-induced bias to later correct it.

\textbf{Conditional Value-at-Risk:}
The Conditional Value-at-Risk (CVaR) (otherwise known as expected shortfall) is another risk metric used jointly with VaR. The great advantage of CVaR is that it is a coherent risk measure and therefore provides a more principled view on risks associated with a given portfolio~\cite{glasserman2013monte}.
While VaR is defined as the quantile of a loss distribution, CVaR corresponds to a conditional expectation on losses. More precisely, CVaR estimates the expected loss conditioned to the fact that the loss is already above the VaR.
For a level of tolerance $\alpha$, $\text{CVaR}_\alpha$ is defined as follows:
$$
\text{CVaR}_\alpha \left( \text{PnL}(\omega) \right) =
    -E_{\mathbb{P}} \left[ 
        PnL(\omega) | - PnL(\omega) > \text{VaR}_{\alpha}
    \right]
.
$$
The results reported in Figure~\ref{fig:cvar} demonstrate that while the use of mixed precision on TPU introduces little bias (less than $1\%$ in relative magnitude compared to a double precision simulation) in the estimation of CVaR it also comes with substantial speed ups.
As for the computation of VaR, the results indicate that TPUs are therefore good candidates for the use of MLMC to produce unbiased estimates rapidly.

\subsection{Monte Carlo American option pricing}
Monte Carlo option pricing of an American option with multiple underlyings presents the computational difficulties encountered in European Basket option pricing because of the need for the simulation of multi-variate diffusion processes while adding the additional complexity of having to proceed with dynamic programming.

\subsubsection{Longstaff-Schwartz pricing of an American Maximum Option}
\begin{figure}[ht]
    \centering
    \includegraphics[width=1.1\linewidth]{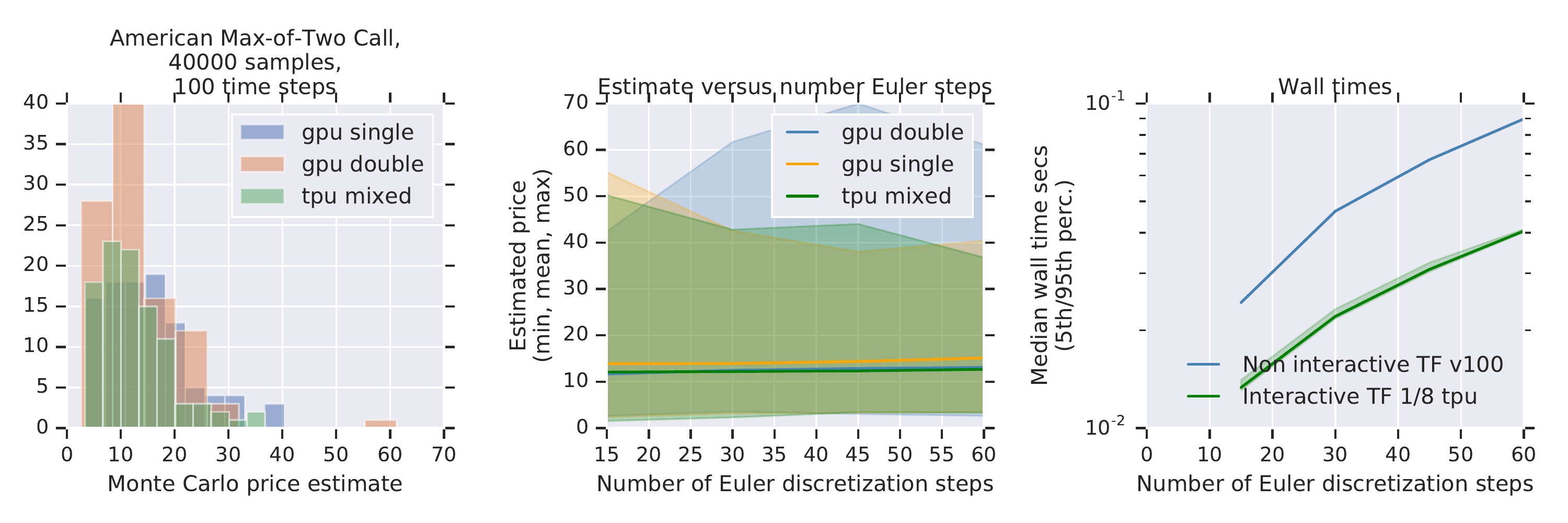}
    \caption{\textbf{Longstaff-Schwartz pricing of an American Maximum Option:} LSM is employed on both TPU and GPU to price an American Max-of-Two Call.
    The very setting of Example 8.6.1 from~\cite{glasserman2013monte} is reproduced and the estimates on all hardware are close to the stated true price of $13.90$.}
    \label{fig:LSM_max_of_two}
\end{figure}

The Longstaff-Schwartz (LSM) method for multi-variate American option pricing relies on a set of samples of the paths of the underlyings to compute the terminal payoff at the expiration of the option assuming there was no early exercise.
As explained earlier in sub-section~\ref{LSM}, LSM then proceeds with dynamic programming taking the form of chained linear regressions by Ordinary Least Squares.
In pratice, we use the Cholesky based solver for such systems integrated in Tensorflow after a Cholesky decomposition of the Grammian.
The MXU is now used for both the forward simulations and the linear inversions in dynamic programming. Therefore, from a numerical standpoint, there is an added level of uncertainty related to the impact of the use of bfloat16 in the MXU.
The mixed precision could be causing numerical instability problems.
In practice, this has not been the case on TPU and we have had to add a regularization term to the regression only to help the GPU implementation which was often failing to do the Cholesky decomposition.
We reproduced the experiment reported in~\cite{glasserman2013monte} in Example 8.6.1 which assesses the true price as $13.90$.
The results presented in Figure~\ref{fig:LSM_max_of_two} show that no significant bias was introduced by the use of TPU while substantial speedups were gained.
It is noteworthy that here we only use one TPU core because there are multiple ways of parallelizing LSM and discussing their different properties is beyond the scope of the present paper.

\subsection{Quasi Monte Carlo Basket Option pricing}\label{sec:sobol}
\begin{figure}[ht]
    \centering
    \includegraphics[width=0.7\linewidth]{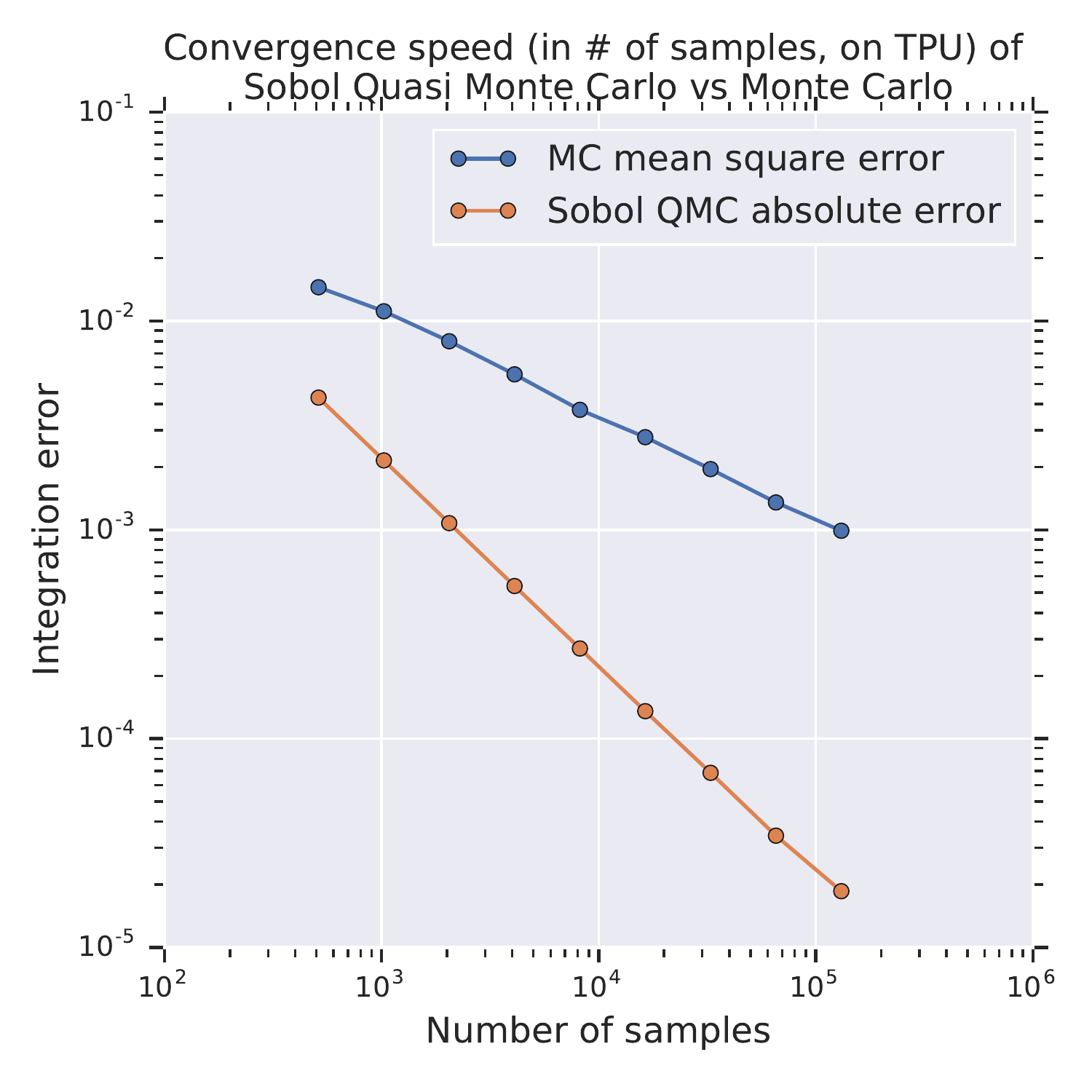}
    \caption{\textbf{Monte Carlo vs Sobol Quasi Monte Carlo} for the pricing of a $16$ asset European Basket Call (we use a single discretization time step) on TPU. While generating Sobol points has a wall time for sampling similar to Threefry, we observe a near linear rate of convergence for QMC as opposed to square root for MC.}
    \label{fig:sobol}
\end{figure}

Quasi Monte Carlo methods considerably speed up the convergence rate of stochastic integration.
Here we consider a workload consisting in simulating the payoff of a European Basket option averaging $16$ correlated assets whose correlations are estimated from historical values as in~\ref{sec:CVAR} (we take the first 16 underlying from 2048). 
The setting we consider is therefore very similar to the numerical examples in 5.5.1 of~\cite{glasserman2013monte} although we consider an arithmetic average basket call as opposed to a geometric average.
A converged Monte Carlo method estimates a price of $6.956624$ ($3.31 \times 10^{11}$ samples) in single precision.
We compare the execution speed on TPU of a standard Monte Carlo simulation and a Quasi Monte Carlo simulation relying on the Sobol sequence.
We only use a single discretization step while sampling. The interest rate, maturity and strike are kept identical to~\ref{sec:basket_option}. Therefore we sample from a $16$ dimensional log-normal distribution prior to computing the discounted payoff.
As we employ the Box-Muller algorithm to generate random samples from uniform samples, we consider the Sobol sequence in $32$ dimensions with direction numbers from~\cite{joe2008constructing}.
We observe empirically similar end-to-end median wall-time in the cloud when compared with our Monte Carlo sampler (at equal number of samples). For instance, we measure a median wall time over $100$ trials of $1.15$ms for $8192$ payoff samples with the Sobol sequence and $1.40$ms with the Threefry-based Monte Carlo method.
As expected, in Figure~\ref{fig:sobol}, we observe that the Root Mean Square Error of the Monte Carlo estimates (over $100$ trials with different Threefry key values) decreases at a rate of $O(\frac{1}{\sqrt{N}})$ where $N$ is the number of samples.
In comparison, the absolute error of the Sobol-based Quasi Monte Carlo integration decreases at a near linear rate.

\section{PROGRAMMING SIMULATIONS WITH TENSORFLOW}\label{sec:programming}

\subsection{Minimal code to run a Monte Carlo simulation}
\begin{figure}
    \centering
    \includegraphics[width=\linewidth]{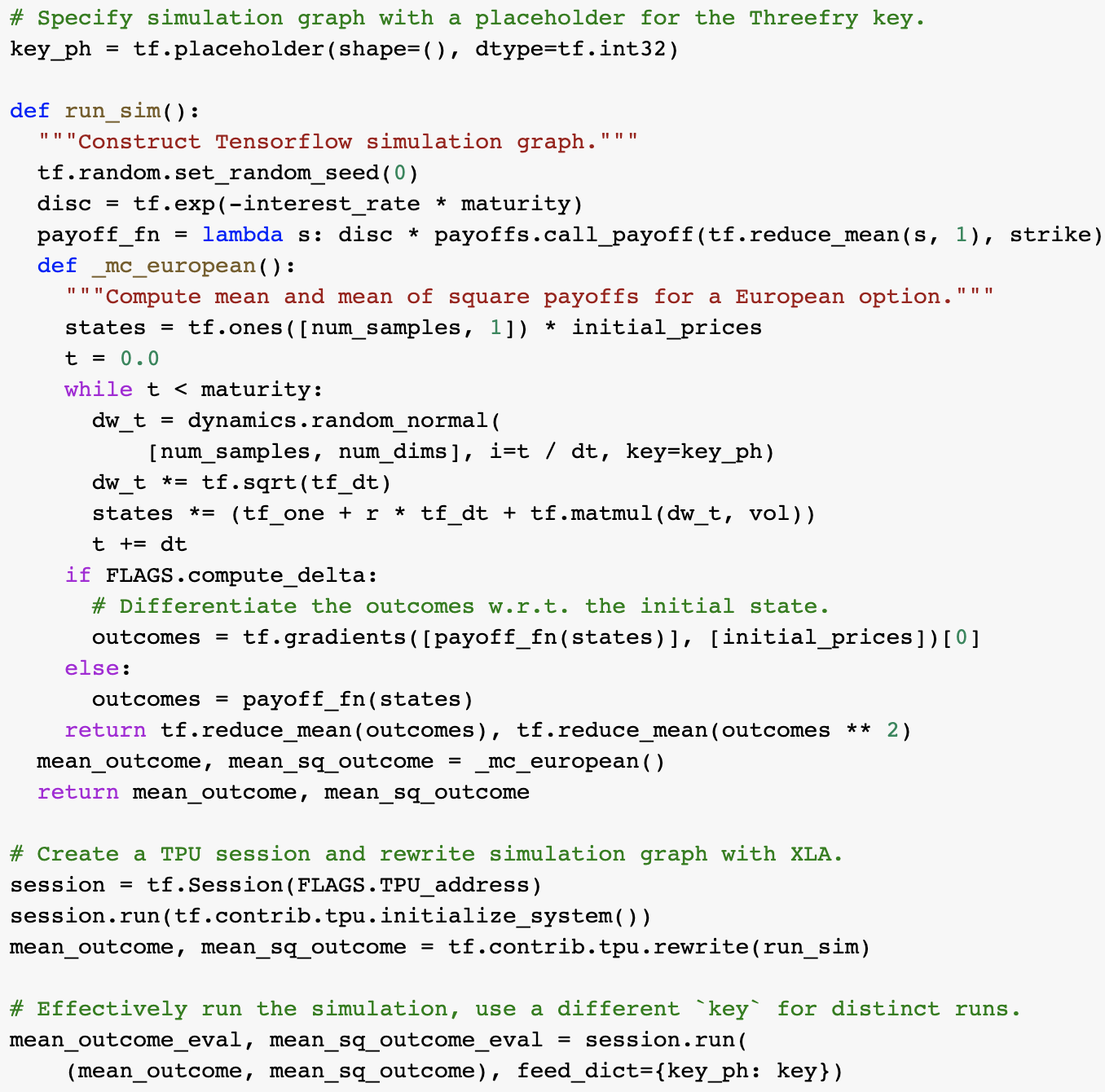}
    \caption{
        Setting up a simulation in an interactive notebook.
        \texttt{dynamics.random\_normal} is a simple wrapper for \texttt{tf.random.stateless\_normal} (i.e. Threefry).
    }
    \label{fig:code_snippet}
\end{figure}

In Figure~\ref{fig:code_snippet}, we show the sufficient code to set up a Monte Carlo simulation in Tensorflow.
One can devise a simulation for a European Basket Option price estimator with a few lines of code, all while working in an interactive Colaboratory (or Jupyter) notebook.
Tensorflow is now a very expressive open source library one can easily extend and therefore provides a lot of flexibility to program various Monte Carlo simulations.
Calling \texttt{tf.contrib.tpu.rewrite} compiles the Tensorflow code with XLA to run optimally on TPU.

\subsection{Automated Differentiation with one line of code}
In Figure~\ref{fig:code_snippet}, we also show that a single line of code suffices to turn payoff estimates into sensitivity estimates (first order derivatives with respect to the initial price in this case computed by AAD, i.e. back-propagation).
This is a remarkable consequence of employing Tensorflow which is optimized for linear algebra acceleration, integrates fast random number generation, and also provides automated direct differentiation.

\section{Conclusion}
In conclusion, we argue that TPUs are indeed accurate enough, fast and easy to use for financial simulation.
Our experiments on multiple workloads demonstrate that even for large simulations written in Tensorflow, TPUs enable a responsive interactive experience with a higher speed than GPUs running Tensorflow.
We showed that Monte Carlo and Quasi-Monte Carlo integration relying on vectorized operations, large matrix/matrix products and linear system inversions is well suited for acceleration on TPU.
Furthermore, programming with Tensorflow makes it trivial to estimate sensitivities with automated differentiation.

Finally, if cases arise in which the mixed precision calculations running on TPUs create too significant a bias, Multi-Level Monte Carlo (MLMC~\cite{giles2008multilevel}) offers an efficient way of correcting this bias while preserving the TPU speed-up.
Therefore, as next steps, we want to experiment with MLMC methods on TPUs with CPUs running a reference precision simulation.
We also plan to use multiple TPUs to scale simulations up.

\newpage

\bibliographystyle{acm}
{\small
\bibliography{main}
}

\end{document}